\documentclass[aps,superscriptaddress]{revtex4-1}

\usepackage[T1]{fontenc}
\usepackage[sc]{mathpazo}
\usepackage{amsmath}
\usepackage{amssymb}
\usepackage{enumerate}
\usepackage{amsthm}

\usepackage{amsfonts,mathrsfs}
\usepackage[style]{fncychap}
\usepackage{graphicx} 
\usepackage{geometry} 
\usepackage{eepic}
\usepackage{ifthen}
\newboolean{ElectronicVersion}
\setboolean{ElectronicVersion}{true} 

\usepackage{hyperref}
\hypersetup{pdfpagemode=UseNone}

\def\be{\begin{equation}}
\def\ee{\end{equation}}
\def\bea{\begin{eqnarray*}}
\def\eea{\end{eqnarray*}}

\newenvironment{mylist}[1]{\begin{list}{}{
    \setlength{\leftmargin}{#1}
    \setlength{\rightmargin}{0mm}
    \setlength{\labelsep}{2mm}
    \setlength{\labelwidth}{8mm}
    \setlength{\itemsep}{0mm}}}
    {\end{list}}

\theoremstyle{definition}

\numberwithin{equation}{section}

\newcounter{questionnumber}

\def\>{\rangle}
\def\<{\langle}

\begin{document}

\title{Nonlocality as a consequence of complementarity}
 \author{Hai Wang}
 \email{3090101669@zju.edu.cn}
 \affiliation{School of Mathematical Sciences, Zhejiang University, Hangzhou 310027, PR~China}
 \author{Asutosh Kumar}
 \email{usashrawan@gmail.com}
 \affiliation{The Institute of Mathematical Sciences, CIT Campus, Taramani, Chennai 600113, India}
 \affiliation{Homi Bhaba National Institute,  Anushaktinagar, Mumbai 400094, India}
 \author{Junde Wu}
 \email{Corresponding author: Junde Wu (wjd@zju.edu.cn)}
 \affiliation{School of Mathematical Sciences, Zhejiang University, Hangzhou 310027, PR~China}


\begin{abstract}
Complementarity and nonlocality are two characteristic traits of quantum physics that distinguishes it from classical physics. In this paper, we prove that the complementarity between global and local observables in Bell's experiment sets the decisive foundation for the nonlocality of composite systems.

\end{abstract}

\maketitle

\section{Introduction}
Quantum theory was conceived, in the early twentieth century, to explain physical phenomena observed at atomic and sub-atomic scales which classical physics could not account for. After a century, it has become a full-fledged fundamental theory of nature at the microscopic level. Since its advent, it has continuously surprised us, by showing to us what can never be expected in the classical domain. Among several seminal developments, complementarity \cite{bohr} and nonlocality \cite{brunner} are eminent concepts. The notion of complementarity has been used in a variety of ways, denoting different concepts and relationships \cite{note1}. Complementarity, in a broad way, refers to a relationship between apparently opposing, contradicting notions or principles, such as the wave and particle nature of light, that together exhaust the possibilities. Bohm refers complementarity to pairs of variables by stating \cite{bohm}: \emph{at the
quantum level, the most general physical properties of any system
must be expressed in terms of complementary pairs of variables, each
of which can be better defined only at the expense of a
corresponding loss in the definition of the other}. In this paper, by complementarity we mean the non-commutativity of two quantum observables, $[X,Y] = XY - YX \neq 0$. For example, in quantum mechanics, position and momentum of a physical system are complementary observables. A pair of complementary observables cannot be observed or measured simultaneously. 
On the other hand, nonlocality--an exotic feature of quantum physics--has proved an indispensable resource for quantum information processing tasks, including communication and computation \cite{nielsen, preskill, wilde}. Phenomena like quantum teleportation \cite{teleportation} and superdense coding \cite{densecoding}, which are not observed in the classical domain, rely
heavily on the nonlocality feature of quantum mechanics. It is understood 
that nonlocality arises between events in subsystems which are not
precisely predictable while at the same time bounded by a
well-defined variable of the whole system which they collectively
constitute \cite{nikolaus}. If events cannot be predicted
precisely in quantum mechanics means there is complementarity among some
observables. This suggests that there is a link between
complementarity and nonlocallity. It is believed that the complementarity between global and local 
observables, in Bell's experiment, leads to nonlocality of composite systems in quantum theory.
The connections between complementarity and 
nonlocality has been explored by several authors \cite{atmanspacher, jonathan, wolk}.
Believing in the idea that nonlocality is a consequence of complementarity, to know how nonlocal a composite
system can be, it is necessary to know to what extent this
complementarity can be. This is the subject matter of this paper.


\section{CHSH Model}
\label{sec-bell}
In this section, we briefly recall the Clauser, Horne, Shimony, and Holt (CHSH) version of Bell inequality \cite{brunner, bell, clauser1, clauser2}.

\subsection{Bell inequality}
Quantum mechanics is a nonlocal theory in the sense that it violates Bell inequality--a mathematical inequality involving certain averages of correlations of measurements, derived using the assumptions of locality and realism. 
That is, quantum mechanics cannot be both local and realistic.
The experimental setting of Bell's test is as follows. There are two observers, Alice (A) and Bob (B). Each of them has two measurement settings: \(A_k\) and \(B_k\), $(k=1,2)$. All these observables are dichotomic, i.e., they take values $\pm 1$. The measurement outcomes of these observables are governed by a joint probability distribution. 
The CHSH version of Bell inequality is expressed as
\begin{equation}
\label{eq:chsh}
|\< A_1B_1 +  A_1B_2 +  A_2B_1 -  A_2B_2 \>| \leq 2,
\end{equation}
where $\< XY \> = \sum_{i,j} x_iy_jp(x_i,y_j)$. This inequality is valid in any physical theory that is local and realistic, and where the physical observables assume the values $\pm 1$.
Now, let \(A_{k}\) and \(B_{k}\) denote the single-qubit Hermitian operators 
\begin{align}
A_k &= a^{(k)}.\sigma = \sum_{i=1}^3 a^{(k)}_i \sigma_i \nonumber \\
B_k &= b^{(k)}.\sigma = \sum_{i=1}^3 b^{(k)}_i \sigma_i, \nonumber 
\end{align}
where $a^{(k)},~b^{(k)}$ are unit vectors in $\mathbb{R}^3$, and $\sigma_i$ are Pauli matrices.
Recall that if Alice and Bob share the singlet, $|\psi^-\rangle_{AB} = \frac{|01\rangle - |10\rangle}{\sqrt{2}}$, then quantum mechanics says that $\< A_kB_{k'} \> =  - a^{(k)}.b^{(k')}$. For the choice of real unit vectors ${a}^{(1)}=(1,0,0),~{a}^{(2)}=(0,0,1),~
{b}^{(1)}=\frac{1}{\sqrt{2}}(1,0,1)$, and ${b}^{(2)}=\frac{1}{\sqrt{2}}(1,0,-1)$, quantum mechanics clearly violates the CHSH inequality. The operator $\mathcal{B} = A_1 \otimes (B_1 + B_2) + A_2 \otimes (B_1 - B_2)$ in CHSH inequality (\ref{eq:chsh}) is called the Bell operator. Note that the Bell operator is a global (or, nonlocal) observable since it is an observable of the whole system. On the other hand, we can define a local observable as $A(r) \otimes B(s) = (\sum_{i=1}^3 r_i \sigma_i) \otimes (\sum_{j=1}^3 s_j \sigma_j)$, where $r$ and $s$ are unit vectors in $\mathbb{R}^3$. The Bell operator, in general, does not commute with the local operator. Then, it is natural to use a norm on the commutator of these observables to quantify how
complementary they are. In this paper, we use the Hilbert-Schmit norm of
operators to quantify it. The Hilbert-Schmit norm is unitary invariant with respect to its argument. 
For the Bell operator ${\cal B}$, the quantity
\begin{equation}
{\cal M}_{{\cal B}}=\sup_{{r},{s}} \parallel[{\cal B},~A(r) \otimes B(s)]\parallel_{2},
\end{equation}
where $r$ and $s$ run over the unit spherical face of
$\mathbb{R}^{3}$, quantifies the maximal ``amplitude'' of complementarity of the Bell
operator with the local observable.

\subsection{Generalized Bell operators for two-qubit case}

In the typical CHSH setting, Alice and Bob seperately measure a spin
along some direction each time. And for a single qubit, nontrivial
observables are only spins. Considering this, 
we define the following operator for two qubit system
\begin{equation}
\label{bell-gen}
{\cal A}=\sum_{i,j=1}^{3}\alpha_{ij}\sigma_{i}\otimes \sigma_{j},
\end{equation}
where $\alpha_{ij} \in \mathbb{R}$ and
$\sigma_{i}$ is the canonical Pauli matrix.  Since this is an extension of the Bell operator, we call it the generalized Bell operator (this can be viewed as a correlation tensor!). 
With this development, we are interested in computing the following quantity
\begin{equation}
{\cal M}=\sup_{{r},{s},\alpha_{ij}}\parallel
[\sum_{i,j=1}^{3}\alpha_{ij}\sigma_{i}\otimes
\sigma_{j},~A(r) \otimes B(s) ]\parallel_{2},
\end{equation}
where $\alpha_{ij}\in\mathbb{R}$ and unit vectors ${r}$, ${s}$ run over all possible choices.
However, we demand ${\cal M}$ to be bounded. For this, we normalize $\alpha_{ij}$:
$\sum_{i,j=1}^{3} \alpha_{ij}^{2}=1$.
Note that due to the Bloch ball structure, for any single-qubit spin operator $A(r) = \sum_{i=1}^3 r_i \sigma_i$, we can always find a unitary operator \(U\) such that $A(r) = U\sigma_3 U$. 
Then for any operator ${\cal A}$ on the two-qubit system, and single-qubit spin operators \(A(r)\) and \(B(s)\), we can find single-qubit unitary operators \(U\) and \(V\) such that
\begin{align}
\parallel [{\cal A}, A(r) \otimes B(s)]\parallel_{2} &= \parallel [{\cal A},U\otimes
V(\sigma_{3}\otimes\sigma_{3})U^{\dagger}\otimes V^{\dagger}]\parallel_{2} \nonumber \\
&=\parallel U^{\dagger}\otimes V^{\dagger} [{\cal A},U\otimes V(\sigma_{3}\otimes\sigma_{3})U^{\dagger}\otimes
V^{\dagger}]U\otimes V\parallel_{2} \nonumber \\
&=\parallel[U^{\dagger}\otimes V^{\dagger} {\cal A} U\otimes V, \sigma_{3}\otimes\sigma_{3}]\parallel_{2}.
\end{align}
Let ${\cal A}$ be the generalized Bell operator (\ref{bell-gen}). 
Since Pauli matrices $\{\sigma_{i}\}_{i=1}^{3}$
are traceless, hermitian and orthogonal with
$tr(\sigma_{i}\sigma_{j})=2\delta_{ij}$, therefore $\{U\sigma_{i}U^{\dagger}\}_{i=1}^{3}$ are also traceless, hermitian and orthogonal. Furthermore, when
$$ U\sigma_{i}U^{\dagger}=\sum_{j=1}^{3}r^{(i)}_{j}\sigma_{j},~~(i=1,2,3)$$
then $\{{r}^{(i)}\}_{i=1}^{3}$ forms an orthonormal basis of $\mathbb{R}^{3}$. 
Hence, for the generalized Bell operator (\ref{bell-gen}), we have
\begin{align}
U\otimes V {\cal A} U^{\dagger}\otimes V^{\dagger} &=U\otimes
V \left(\sum_{i,j=1}^{3}\alpha_{ij}\sigma_{i}\otimes\sigma_{j} \right) U^{\dagger}\otimes V^{\dagger} \nonumber \\
&=\sum_{i,j=1}^{3}\alpha_{ij} \left(\sum_{p=1}^{3} r^{(i)}_{p} \sigma_{p} \right) \otimes \left(\sum_{q=1}^{3} s^{(j)}_{q} \sigma_{q}\right) \nonumber \\
&=\sum_{p,q=1}^{3} \left(\sum_{i,j=1}^{3} \alpha_{ij}r^{(i)}_{p}s^{(j)}_{q} \right) \sigma_{p}\otimes \sigma_{q}, \nonumber \\
&= \sum_{p,q=1}^{3} \alpha_{pq} \sigma_{p}\otimes \sigma_{q}, 
\end{align}
where we write the last step from the orthonormality of bases $\{{r}^{(i)}\}_{i=1}^{3}$ and
$\{{s}^{(j)}\}_{j=1}^3$, and the normalization
\begin{equation}
\sum_{p,q=1}^{3} \left(\sum_{i,j=1}^{3} \alpha_{ij}r^{(i)}_{p}s^{(j)}_{q} \right)^2 = \sum_{i,j=1}^{3}\alpha_{ij}^{2} \sum_{p=1}^3 \left(r^{(i)}_{p}\right)^{2} \sum_{q=1}^3 \left(s^{(j)}_{q} \right)^{2} = \sum_{i,j=1}^{3}\alpha_{ij}^{2}=1.
\end{equation}
Now, since $[\sum_{i,j=1}^3 \alpha_{ij}\sigma_{i}\otimes \sigma_{j},\sigma_{3}\otimes \sigma_{3}]=2i(\alpha_{23}\sigma_{1}\otimes I+\alpha_{32}I\otimes \sigma_{1}-\alpha_{13}\sigma_{2}\otimes I-\alpha_{31}I\otimes \sigma_{2})$ and $\sum_{i,j=1}^3 \alpha_{ij}^{2}=1$, we have
\begin{align}
{\cal M} &=\sup_{{r},{s},\alpha_{ij}}\parallel [\sum_{i,j=1}^{3}\alpha_{ij}\sigma_{i}\otimes \sigma_{j},~A(r) \otimes B(s) ]\parallel_{2} \nonumber \\
&= \sup_{\alpha_{pq}} \parallel [\sum_{p,q=1}^{3}\alpha_{pq}\sigma_{p}\otimes \sigma_{q},~\sigma_3 \otimes \sigma_3 ]\parallel_{2} \nonumber \\
&= \sup_{\alpha_{pq}} 4\sqrt{\alpha_{13}^2 + \alpha_{23}^2 + \alpha_{31}^2 + \alpha_{32}^2} \leq 4.
\end{align}
This upper bound is saturated by the classic Bell
operator $${\cal B}_{0}=\sigma_{1}\otimes
\left(\frac{\sigma_{1}+\sigma_{3}}{\sqrt{2}}+\frac{\sigma_{1}-\sigma_{3}}{\sqrt{2}}\right)+\sigma_{3}\otimes
\left(\frac{\sigma_{1}+\sigma_{3}}{\sqrt{2}}-\frac{\sigma_{1}-\sigma_{3}}{\sqrt{2}}\right)=\sqrt{2}(\sigma_{1}\otimes
\sigma_{1}+\sigma_{3}\otimes \sigma_{3}).$$
The normalized form of
${\cal B}_{0}$ is ${\cal B}_{0}'=\frac{\sigma_{1}\otimes
\sigma_{1}+\sigma_{3}\otimes \sigma_{3}}{\sqrt{2}}$, for which we can show that ${\cal M} = \sup_{{r}, {s}}\parallel
[{\cal B}_{0}', A(r)\otimes B(s)]\parallel_{2}=4$.
Next, let us consider the extreme case: ${\cal M}=0$. This means that we can always measure global observables $\sum_{i,j=1}^{3}\alpha_{ij}\sigma_{i}\otimes\sigma_{j}$ and local observables $A(r)\otimes B(s)$ simultaneously. Especially, for a pure state in such a theory, the Bell operator $\mathcal{B} = A_1 \otimes (B_1 + B_2) + A_2 \otimes (B_1 - B_2)$ and local operators $A_k \otimes B_{k'}~~(k,k'=1,2)$ should have exact values together. 
Since $A_k$ and $B_k$ are dichotomic, i.e., they take values $\pm 1$, the allowed values of ${\cal B}$ lie in $[-2,+2]$. So a theory, in which ${\cal M}=0$, is local in terms of Bell inequalities.


\subsection{Two-qudit case and the generalized Pauli matrices}
In case of two-qudit systems, we can carry out the above treatment in terms of generalized Pauli matrices. 
In this paper, we consider the (discrete) Weyl operators as the generalized Pauli matrices. 
Formally, the Weyl operator basis in \(d\)-dimensional complex space is given by the following set \cite{watrous}:
$$\widetilde{\sigma} = \{X^{a}Z^{b}:(a,b)\in \mathbb{Z}^{2}_{d}\},$$
where $X=\sum_{a\in\mathbb{Z}_{d}}E_{a+1,a}$ and $Z=\sum_{a\in\mathbb{Z}_{d}}\omega^{a}E_{a,a}$ are unitary operators with $\omega=\exp(2\pi i/d)$, and $\mathbb{Z}_{d}=\{0,1,...,d-1\}$ is a {\em ring} with respect to addition and multiplication defined over modulo \(d\).
It should be noted here that though we can have generalized Bloch vectors in higher dimensions corresponding to the Weyl operators, these generalized Bloch vectors do not together form the structure of a ball as in the two-dimensional case. Moreover, since the Weyl basis does not consist of hermitian operators, these generalized Bloch vectors are located in a complex space. 
Below, for the sake of completeness, we recall a few basic properties of the Weyl operators:
(i) $tr(X^{a}Z^{b})=d$ if $a=b=0$, and zero otherwise. (ii) The above collection forms an orthogonal basis because $\langle X^{a}Z^{b},X^{c}Z^{d}\rangle=tr(Z^{-b}X^{-a}X^{c}Z^{d})=tr(X^{c-a}Z^{d-b})=d\delta_{(a,b),(c,d)}$. (iii) \(X\) and \(Z\) obey the commutation relation, $ZX=\omega XZ$, which (for instance) implies $(X^{a}Z^{b})(X^{c}Z^{d})=\omega^{bc}X^{a+c}Z^{b+d}=\omega^{bc-ad}(X^{c}Z^{d})(X^{a}Z^{b})$.\\

Using the above commutation relation, for every $a,b\in \mathbb{Z}_{d}$, we will get $$X^{a}Z^{b}Z=X^{a}Z^{b+1},\ ZX^{a}Z^{b}=\omega^{a}X^{a}Z^{b+1}.$$
Besides, for indices $(a,b)$, we can define $k=ad+b$. As a result of this, we can put members of the Weyl basis by the order of their indices. For example, we can use $\widetilde{\sigma}_{1}$ to represent $X^{0}Z^{1} = Z$. Similarly,
\begin{equation}
X^{a}Z^{b}Z=\widetilde{\sigma}_{k}\widetilde{\sigma}_{1}=X^{a}Z^{b+1}=\widetilde{\sigma}_{k+1},
\end{equation}
where $k=ad+b$, and $k+1$ corresponds to indices $(a,b+1)$ under the addition of direct product of $\mathbb{Z}_{d}$. We denote the identity operator in the Weyl basis by $\widetilde{\sigma}_{0}$.\\

For two-qudit systems, as two-qubit systems, we can define global observables $\sum_{i,j=1}^{d^{2}-1}\alpha_{ij}\widetilde{\sigma}_{i}\otimes \widetilde{\sigma}_{j}$, local observables $\widetilde{A}(r) \otimes \widetilde{B}(s)$, and the quantity
\begin{align}
{\cal M}_{d} &:= \sup_{r,s,\alpha_{ij}} \parallel [\sum_{i,j=1}^{d^{2}-1}\alpha_{ij}\widetilde{\sigma}_{i}\otimes \widetilde{\sigma}_{j}, \widetilde{A}(r) \otimes \widetilde{B}(s)] \parallel_{2} \nonumber \\
&:= \sup_{\alpha_{ij},U,V}\parallel [\sum_{i,j=1}^{d^{2}-1}\alpha_{ij}\widetilde{\sigma}_{i}\otimes \widetilde{\sigma}_{j}, U\otimes V(\widetilde{\sigma}_{i_{0}}\otimes \widetilde{\sigma}_{i_{0}})U^{\dagger}\otimes V^{\dagger}]\parallel_{2} \nonumber \\
&= \sup_{\alpha_{ij},U,V}\parallel [U\otimes V \left(\sum_{i,j=1}^{d^{2}-1}\alpha_{ij}\widetilde{\sigma}_{i}\otimes \widetilde{\sigma}_{j} \right) U^{\dagger}\otimes V^{\dagger}, \widetilde{\sigma}_{i_{0}}\otimes \widetilde{\sigma}_{i_{0}}]\parallel_{2},
\end{align}
where \(U\), \(V\) are unitary operators on $\mathbb{C}^{d}$, and $\widetilde{\sigma}_{i_{0}}$ is one of the elements of the Weyl basis (excluding the identity operator). Now, an important question is how to compute ${\cal M}_d$. Recall that the Weyl operators are traceless and orthogonal 
with $\langle A,B\rangle=tr(A^{\dagger}B)=d\delta_{AB}$.
So for any unitary operator $U$, the set
$\{U\widetilde{\sigma}_{i}U^{\dagger}\}_{i=1}^{d^2-1}$ is still orthogonal and
traceless. That is, for every $i$,$$U\widetilde{\sigma}_{i}U^{\dagger}=\sum_{p=1}^{d^{2}-1}\mu^{(i)}_{p}\widetilde{\sigma}_{p},$$
where $\mu^{(i)}$ is a $(d^2-1)$-dimensional vector, and the set $\{{\mu}^{(i)}\}_{i=1}^{d^{2}-1}$
constitutes an orthonormal basis of $\mathbb{C}^{d^{2}-1}$. Therefore,
\begin{equation}
U\otimes V \left(\sum_{i,j=1}^{d^{2}-1}\alpha_{ij}\widetilde{\sigma}_{i}\otimes \widetilde{\sigma}_{j} \right) U^{\dagger}\otimes V^{\dagger}=\sum_{i,j,p,q=1}^{d^{2}-1}\alpha_{ij}\mu^{(i)}_{p}\nu^{(j)}_{q}\widetilde{\sigma}_{p}\otimes \widetilde{\sigma}_{q}
\end{equation}
where the vector sets $\{\textbf{$\mu$}^{(i)}\}_{i=1}^{d^{2}-1}$ and $\{\textbf{$\nu$}^{(j)}\}_{j=1}^{d^{2}-1}$ are two orthonormal bases of $\mathbb{C}^{d^{2}-1}$. Again, because of the orthonormality of the two bases, and if we demand that $\sum_{i,j=1}^{d^{2}-1}\mid \alpha_{ij}\mid^{2}=1$, then we can have
$$\sum_{p,q=1}^{d^{2}-1}\mid \sum_{i,j=1}^{d^{2}-1}\alpha_{ij}\mu^{(i)}_{p}\nu^{(j)}_{q}\mid^{2}=1.$$
Here, unlike two-qubit case, we consider $\alpha_{ij}$ to be complex. This is because even if we assume that $\alpha_{ij}$ are real, this does not ensure that $\sum_{i,j=1}^{d^{2}-1}\alpha_{ij}\mu^{(i)}_{p}\nu^{(j)}_{q}$ are also real.  
Now, with the notation set earlier for labelling the Weyl operators, we have
$$[\sum_{i,j=1}^{d^{2}-1}\alpha_{ij}\widetilde{\sigma}_{i}\otimes\widetilde{\sigma}_{j},
\widetilde{\sigma}_{1}\otimes
\widetilde{\sigma}_{1}] = \sum_{i,j=1}^{d^{2}-1}\alpha_{ij} \left(1-\omega^{[\frac{i}{d}]+[\frac{j}{d}]} \right) \widetilde{\sigma}_{i+1}\otimes\widetilde{\sigma}_{j+1},$$
where$[\frac{i}{d}]$ and $[\frac{j}{d}]$ represent the integer parts
of $i$ and $j$ divided by $d$. Remember that if $\widetilde{\sigma}_0 = \widetilde{\sigma}_{d^{2}}$ be the identity operator, then there is no repeated terms in $\{\widetilde{\sigma}_{i+1}\otimes\widetilde{\sigma}_{j+1}\}_{i,j=1}^{d^{2}-1}$.
Consequently, for the two-qudit case, the quantity ${\cal M}_d$ is given by
\begin{align}
{\cal M}_d &= \sup_{\alpha_{ij}}\parallel [\sum_{i,j=1}^{d^{2}-1}\alpha_{ij}\widetilde{\sigma}_{i}\otimes \widetilde{\sigma}_{j}, \widetilde{\sigma}_{i_{0}}\otimes \widetilde{\sigma}_{i_{0}}]\parallel_{2} \nonumber \\
&= \sup_{\alpha_{ij}}\parallel\sum_{i,j=1}^{d^{2}-1}\alpha_{ij} \left(1-\omega^{[\frac{i}{d}]+[\frac{j}{d}]} \right) \widetilde{\sigma}_{i+1}\otimes\widetilde{\sigma}_{j+1}\parallel_{2}\leq 2d,
\end{align}
where $\alpha_{ij}\in \mathbb{C}$, $\sum_{i,j=1}^{d^{2}-1}\mid \alpha_{ij}\mid^{2}=1$, and the operator $\widetilde{\sigma}_{i_{0}}$ is an element of the Weyl basis on $\mathbb{C}^{d}$ except the identity operator. In obtaining the above bound, we have used the property, $tr(\widetilde{\sigma}_{i}^{\dagger}\widetilde{\sigma}_{j})=d\delta_{ij}$, of the Weyl operators. Note that the above upper bound (i) is independent of our choice of $\widetilde{\sigma}_{i_{0}}$, and (ii) for \(d=2\), we recover the results of the two-qubit case. 

\section{Complementarity and Nonlocality}
As discussed earlier, the complementarity between global and local observables is believed to set decisive conditions for nonlocality of composite systems. That is, the ``strength'' of this complementarity can imply to what extent the composite system can be nonlocal. In the previous section, we have introduced the quantity ${\cal M}_d$ which shows the maximal complementarity between global and local observables for a two-qudit system. In this connection, we ask: does there exist a quantity that measures the maximal nonlocality of a two-qudit system? If it does, and if complementarity determines nonlocality, then we expect some relation between ${\cal M}_d$ and this quantity. For example, these quantities may be some function of the dimension of the system under consideration. In fact, such a quantity already exists in the literature. In \cite{carlos}, author has defined a ``measure'' of nonlocality, ${\cal K}_d$, which quantifies the largest Bell violation in a two-qudit system.
It is shown that ${\cal K}_d \leq 4d$. Thus, we see that there is an apparent relation between ${\cal M}_d$ and ${\cal K}_d$. That is, these quantities are of the same order with respect to the local dimension \(d\).     
This means that by knowing the maximal complementarity between global observables and local observables, we can get an estimate of how nonlocal the composite system can be at most. This is in consonance with our expectation. 
Remember that we have observed certain constraints in obtaining ${\cal M}_d \leq 2d$. Still ${\cal M}_d$ serves as a valuable estimator of the nonlocality of a composite system. 
It shows a beautiful connection between nonlocality and complementarity.

\section{Conclusion}

Complementarity and nonlocality are two characteristic traits of quantum mechanics, which have profound implications in foundations of quantum physics and in quantum information theory. In this paper, we have introduced the notion of generalized Bell operator, and have defined the quantity ${\cal M}_d$ to show the maximal complementarity between global and local observables. And by the result of \cite{carlos}, we see an apparent relation between complementarity and nonlocality of a composite system, with respect to the dimension of the local system. We believe that our results will shed further light on the study of nonlocality. 

\subsection{ACKNOWLEDGMENTS}

This project is supported by the National Natural Science Foundation
of China (Grants No. 11171301 and No. 11571307) and by the Doctoral
Programs Foundation of the Ministry of Education of China (Grant No.
J20130061). AK acknowledges research fellowship from the Department
of Atomic Energy, Government of India.


\end{document}